\documentclass{iau}
\usepackage{graphicx}

\title[IAUS 302: Surface magnetism of cool giant and supergiant stars] 
{Surface magnetism of cool giant and supergiant stars}

\author[Heidi Korhonen]   
{Heidi Korhonen$^1$}

\affiliation{$^1$Finnish Centre for Astronomy with ESO (FINCA), University of 
  Turku, V{\"a}is{\"a}l{\"a}ntie 20, FI-21500 Piikki{\"o}, Finland\\ 
  email: {\tt heidi.h.korhonen@utu.fi} }

\pubyear{2008}
\volume{xxx}  
\pagerange{119--126}
\jname{Title of your IAU Symposium}
\editors{A.C. Editor, B.D. Editor \& C.E. Editor, eds.}
\begin{document}

\maketitle

\begin{abstract}
The existence of starspots on late-type giant stars in close binary systems, that exhibit rapid rotation due to tidal locking, has been known for more than five decades. Photometric monitoring spanning decades has allowed studying the long-term magnetic activity in these stars revealing complicated activity cycles. The development of observing and analysis techniques that has occurred during the past two decades has also enabled us to study the detailed starspot and magnetic field configurations on these active giants. In the recent years magnetic fields have also been detected on slowly rotating giants and supergiant stars. In this paper I review what is known of the surface magnetism in the cool giant and supergiant stars.
\end{abstract}

\firstsection 
\section{Introduction}

A slightly over a century ago \cite[Hale (1908)]{Hale1908} discovered that sunspots are caused by strong magnetic fields in the Sun. Naturally, observing solar-like magnetic fields in other stars is extremely demanding, and their detection had to wait almost a century. Still, stronger fields could be detected already much earlier. The first detections were done in magnetic, chemically peculiar, Ap stars, which have strongest known magnetic fields of non-degenerate stars (see, e.g., \cite[Babcock 1947a]{Babcock47a}; \cite[Babcock 1947b]{Babcock47b}). Recent years magnetic fields have been discovered in many different classes of stars, from pre-main sequence stars (e.g., \cite[Guenther \etal\ 1999]{Guenther99}) to neutron stars (e.g., \cite[Kouveliotou \etal\ 1998]{Kouveliotou98}), and throughout the main sequence from M-type (e.g., \cite[Donati \etal\ 2006]{Donati06}; \cite[Morin \etal\ 2008]{Morin08}) to O-type (e.g., \cite[Hubrig \etal\ 2008]{Hubrig08}; \cite[Grunhut \etal\ 2009]{Grunhut09}) stars. 

In this review the surface magnetism of cool G--M giants and supergiants is discussed. The main emphasis is on the observations of the surface features caused by magnetic fields and the measurement of the magnetic fields themselves, but also the origins of these fields are discussed.

\section{Methods for studying stellar activity and magnetism}

Obtaining spatial information of the surface of stars, which appear as point sources in our telescopes, is challenging. The two main methods for achieving this are photometry and high resolution spectroscopy through Doppler imaging techniques. Photometry is the easiest and least observing time consuming way of carrying out starspot studies. In many active stars the starspots are so large that they cause brightness variations which can be few tens of percent from the mean light level, thus making them easily observable even from the ground (see, e.g., \cite[Chugainov 1966]{Chugainov66}; \cite[Montle \& Hall 1972]{Montle72}). On the other hand, solar-like spots are so small that they are lost in the noise, unless extremely precise observations can be obtained. Currently the by-far best precision and time resolution observations are provided by the Kepler satellite. 

Doppler imaging (see, e.g., \cite[Vogt \etal\ 1987]{Vogt87}; \cite[Rice \etal\ 1989]{Rice89}; \cite [Piskunov \etal\ 1990]{Piskunov90}), which provides the best spatial resolution on the stellar surface, is a method that uses high resolution, high signal-to-noise spectroscopic observations at different rotational phases of the star. If the star has a non-uniform surface temperature, i.e., has starspots, the spectral lines show small distortions from the normal Gaussian shape. These distortions move in the line-profile when the position of the starspots on the surface changes, due to the change of line-of-sight velocity caused by the stellar rotation. Surface maps, or Doppler images, are constructed by tracking the movement of these distortions by combining all the observations from different rotational phases and typically comparing them with synthetic model line profiles. 

During just the last years a breakthrough using long baseline infrared and optical interferometers has occurred. These facilities now routinely produce aperture synthesis images with milli-arcsecond angular resolution. A variety of targets have been imaged with astonishing result, e.g., bulging stars rotating near their critical limit (\cite[Monnier \etal\ 2007]{Monnier07}) and compact dust disk around a massive young stellar object (\cite[Kraus \etal\ 2010]{Kraus10}). Infrared interferometric imaging has produced amazing results of stellar surfaces and the time is drawing near when even temperature spots on cool stars can be imaged (see \cite[Roettenbacher \etal\ 2013]{Roettenbacher13}), and naturally giant stars are the most fruitful starting point for this due to their large size.

Studying the magnetic fields directly in stars can be done using Zeeman splitting of the spectral lines. This method, which is the easiest one to interpret, requires high spectral resolution and high signal-to-noise ratio to accurately observe the spectral-line profile shapes (see e.g., \cite[Johns-Krull \& Valenti 1996]{JohnsKrull96}). Another way to study the magnetic fields themselves is to measure the polarisation of spectral lines (e.g., \cite[Donati \etal\ 1997]{Donati97}). In astronomy polarisation is usually described using the Stokes vector [I, Q, U, V], where Stokes I gives the total intensity spectrum, Stokes V describes the circular polarisation and Stokes Q and U the linear polarisation. Circular polarisation is mainly sensitive to the line-of-sight magnetic field and the linear one to the perpendicular field, therefore polarisation can be also used to measure the orientation of the field. If spectropolarimetric observations are obtained over the stellar rotation, similarly to what is done in Doppler imaging, the surface magnetic field of the star can mapped using Zeeman-Doppler imaging technique (\cite[Semel 1989]{Semel89}).

\section{Active giants}

Cool giant stars are in general very slow rotators. Many of their progenitors were slow rotators when in the main sequence and, even if they were not, the expanding envelope reduces the rotation rate significantly. This makes rotation driven, solar-like, dynamo action very improbable in these stars. Still, there are some cases of giant stars being rapid rotators, and magnetically very active. Reviews of starspots and their properties, also in active giants, are given by \cite[Strassmeier (2009)]{Strassmeier09} and \cite[Berdyugina (2005)]{Berdyugina05}. The main types of active giants are discussed in the following.

\subsection{RS CVn-type binaries}

\begin{figure}[t]
\begin{center}
 \includegraphics[width=3.4in]{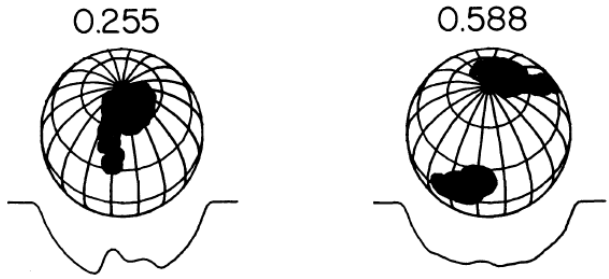} 
 \caption{Examples of first Doppler maps of HR1099. (from \cite[Vogt \& Penrod 1983]{Vogt83})}
   \label{fig1}
\end{center}
\end{figure}

The most common type of active giants is the RS~Cvn-type binaries, where a G--K giant or sub-giant is partnered with a less massive G--M main sequence star or a sub-giant. These typically tidally locked binaries are still detached systems, and their orbital (and therefore often also rotation periods) are commonly from few days to some tens of days. General properties and classification of RS~CVn-type binaries are discussed, e.g., by \cite[Hall (1976)]{Hall76} and \cite[Morgan \& Eggleton (1979)]{Morgan79}. The first star on which starspots were suggested to be the cause of the brightness variations, AR Lac (\cite[Kron 1947]{Kron1947}), is an RS~Cvn-type binary. As is also the first cool star for which Doppler imaging technique was applied, HR1099, shown in Fig.\,\ref{fig1} (\cite[Vogt \& Penrod 1983]{Vogt83}). 

\begin{figure}[t]
\begin{center}
 \includegraphics[width=3.4in]{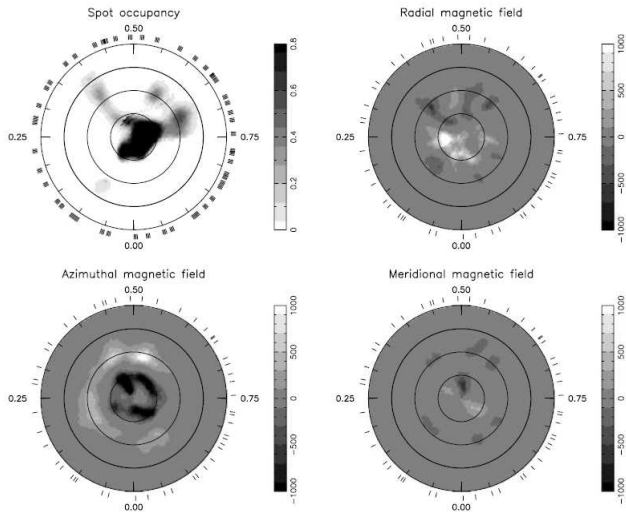} 
 \caption{Zeeman Doppler maps of HR 1099. The flattened polar view in the upper-left panel corresponds to a brightness image and the other panels show radial, azimuthal and meridional components of the field in the upper-right, lower-left and lower-right panels, respectively. (from \cite[Petit \etal\ 2004a]{Petit04a})}
   \label{fig2}
\end{center}
\end{figure}

The Doppler imaging of RS~CVn binaries often shows very large starspot, like the 12$\times$20 solar radii spot on HD~12545 (\cite[Strassmeier 1999]{Strassmeier99}), and also very high spot latitudes, or even polar spots (e.g., in KU~Peg by  \cite[Weber \& Strassmeier 2001]{Weber01}). For a handful of RS~CVn-type binaries magnetic field maps have been obtained, among these stars are HR~1099 and II~Peg which have maps from multiple epochs. And example of such a maps is shown in Fig.\,\ref{fig2}.  \cite[Donati (1999)]{Donati99}, \cite[Donati \etal\ (2003)]{Donati03}, and \cite[Petit \etal\ (2004a)]{Petit04a} have obtained both spot filling-factor maps and magnetic field maps of HR~1099 on five epochs spanning more than a decade in total. They discovered two distinct azimuthal field regions of opposite polarity around latitudes 30$^{\circ}$ and 60$^{\circ}$. They also conclude that small-scale brightness and magnetic features undergo changes at time scales of 4--6 weeks, whereas the large-scale structures are stable over several years. Recently, \cite[Kochukhov \etal\ (2013)]{Kochukhov13} published magnetic field and temperature maps of II~Peg obtained from seven different epochs spanning in total 12 years. They observe significant field evolution on the time scale of their observations, additionally they did not find a clear correlation between magnetic and temperature features in their maps.

\subsection{FK Comae-type giants}

FK~Comae-type giants are a very small group of highly active G--K giants and subgiants (\cite[Bopp \& Rucinski 1981]{BoppRucinski}; \cite[Bopp \& Stencel 1981]{BoppStencel}). They show activity levels similar to RS~CVn binaries. What makes them special is that they are extremely rapidly rotating (FK~Com itself has $v\sin i$ around 160km/s), but still they are not part of a binary systems. The most commonly accepted explanation for the rapid rotation is that they are end products of a coalescence of W~Uma-type contact binaries (e.g., \cite[Bopp \& Rucinski 1981]{BoppRucinski}; \cite[Bopp \& Stencel 1981]{BoppStencel}). 

Temperature maps of three FK~Comae-type stars have been published in the literature. The first one mapped was HD~32918 (YY~Men) by \cite[Piskunov \etal\ 1990]{Piskunov90}. The map shows mainly equatorial spot configurations. The other two FK~Comae stars with temperature maps are FK~Com itself (e.g., \cite[Korhonen \etal\ 1999]{Korhonen99}; \cite[Korhonen \etal\ 2007]{Korhonen07}) and HD~199178 (e.g., Hackman \etal\ 2001). These stars mainly show high latitude spot configurations, but no real polar spots.

Magnetic field of HD~199178 has been mapped using Zeeman Doppler imaging by \cite[Petit \etal\ (2004b)]{Petit04b}. Their maps from 2002 and 2003 reveal large regions of azimuthal field, and also changes in the exact field configuration with a time scale of about two weeks. For FK~Com no Zeeman Doppler map exists, but its mean longitudinal magnetic field has been studied using low resolution spectra (\cite[Korhonen \etal\ 2009]{Korhonen09}). These observations indicate that the two high-latitude spots seen in the contemporaneous temperature map could have different polarity.

\subsection{Lithium rich giants}

Some cool giant stars show enhanced Li abundance, and some of these stars are also moderately rapid rotators, and thus magnetically active. \cite[Fekel \& Balachandran (1993)]{Fekel93} suggested that during the first dredge-up both angular momentum and Li-rich material could be dredged-up from the stellar interior, creating a rapidly rotating Li-rich giant. Further analysis by \cite[Charbonnel \& Balachandran (2000)]{Charbonnel00} of the Li abundance and rotation in giants did not show clear correlation.

Whatever the cause for the rapid rotation is, it still enhances the dynamo operation in these stars and they become magnetically active. Recently, \cite[K{\H o}v{\'a}ri \etal\ (2013)]{Kovari13} published temperature maps of two Li-rich giants, DP~CVn and DI~Psc. For both stars low latitude spots with relatively small temperature contrasts (600-800 K below the unspotted surface temperature) are recovered. In addition higher latitude spots are recovered, but they either have lower contrast (DP~CVn) or smaller extent (DI~Psc). \cite[L{\`e}bre \etal\ (2009)]{Lebre09} also detected Stokes V signal with temporal variations on Li-rich giant HD~232862, indicating a presence of magnetic field in this star.

\section{Slowly rotating giants}

One would not expect to be able to detect magnetic fields on slowly rotating giant stars, because the slow rotation would not enable the dynamo to create strong enough magnetic fields. Still, magnetic fields have been detected, and even mapped, in some slowly rotating giants.

Pollux is a K0 giant with rotation period of 100--500 days. The star has been known for some time to be weakly active (see, e.g., \cite[Strassmeier \etal\ 1990a]{Strassmeier90a}). A weak longitudinal magnetic field of -0.46$\pm$0.04~G was detected in Pollux by \cite[Auri{\`e}re \etal\ (2009)]{Auriere09}, see Fig.\,\ref{fig2}. Another similar case is Arcturus where \cite[Sennhauser \& Berdyugina (2011)]{Sennhauser11} detected a weak longitudinal magnetic field of 0.65$\pm$0.26~G and 0.43$\pm$0.16~G from two spectra obtained at different times. The origin of the field detected in Pollux and Arcturus is most likely solar-like $\alpha\Omega$-dynamo.

\begin{figure}[t]
\begin{center}
 \includegraphics[width=3.4in]{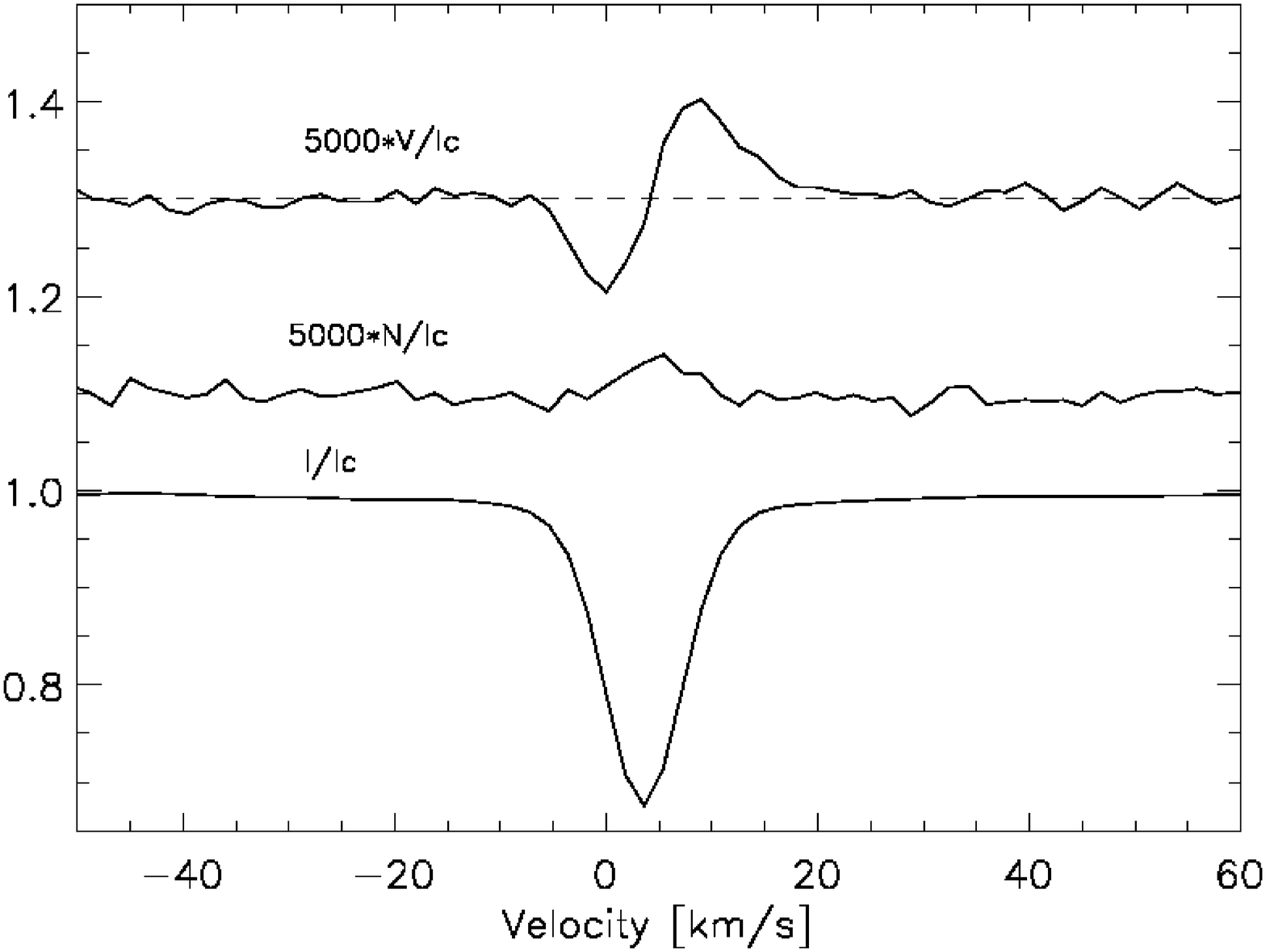} 
 \caption{Stokes V, the diagnostic null profile and the Stokes I observation of Pollux. (from \cite[Auri{\`e}re \etal\ 2009]{Auriere09})}
   \label{fig3}
\end{center}
\end{figure}

Not only weak fields are detected in slowly rotating giants. One puzzling case is EK~Eri, which has rotation period of approximately 300 days (e.g., \cite[Strassmeier \etal\ 1990b]{Strassmeier90b}), and still longitudinal magnetic field as strong as 100~G (\cite[Auri{\`e}re \etal\ 2008]{Auriere08}). The modelling of \cite[Auri{\`e}re \etal\ (2008)]{Auriere08} also reveal that the large scale magnetic field is most likely dominated by a poloidal component. They suggest that EK~Eri is a descendant of a strongly magnetic Ap star. Similar suggestion has been made to explain the magnetic field measurements of $\beta$~Cet. \cite[Tsvetkova \etal\ (2013)]{Tsvetkova13} publish two Zeeman Doppler images of $\beta$~Cet, obtained one year apart and showing very similar magnetic field configurations dominated by a dipole, and very little evolution between the maps. This all implies magnetic Ap star origin of $\beta$~Cet.

Magnetic fields have also been detected in more evolved M giants. EK~Boo is a rapidly rotating (for its class) M5 giant. \cite[Konstantinova-Antova \etal\ (2010)]{KonstantinovaAntova10} detected magnetic field which varied between -0.1~G and -8~G with time. In one of their observations they also detected complex structure in the Stokes~V profile, which could indicate dynamo origin of the field. In their sample of nine M giants \cite[Konstantinova-Antova \etal\ (2010)]{KonstantinovaAntova10} also obtained marginal detection of magnetic field in another star, $\beta$~And.

\section{Supergiants}

If cool giants are typically slow rotators, then M supergiants are even slower rotators. This makes solar-like dynamo operation in them virtually impossible. In addition, due to the very extended envelope of these stars, also a fossil magnetic field, remnant from a magnetic main sequence progenitor, would be extremely diluted and is not expected. 

M supergiants have gigantic convection cells, which have been predicted in simulations (e.g., \cite[Freytag \etal\ 2002]{Freytag02}) and seen in observations of Betelgeuse (e.g., \cite[Gilliland \& Dupree 1996]{Gilliland96}; \cite[Haubois et al. 2009]{Haubois09}). It has also been predicted that local dynamo could operate in the convection cells and create global magnetic field in supergiant stars (\cite[Dorch 2004]{Dorch04}). 

The first measurement of the magnetic fields in M supergiants was done for Betelgeuse by \cite[Auri{\`e}re \etal\ (2010)]{Auriere10}. They detected a weak Stokes V signal, and measured longitudinal field of about one Gauss at six different epochs. There was also some evidence of the field strength increasing during the one month that the observations span. The results by \cite[Petit \etal\ (2013)]{Petit13} indicate that the magnetic elements may be concentrated in the sinking components of the convective flows.

\section{Activity cycles} 

The Sun exhibits well established 11-year spot cycle, and a 22-year magnetic cycle. The Mt. Wilson H\&K project (\cite[Wilson 1978]{Wilson78}) has established similar behaviour of the over-all activity in many cool main sequence stars. Unfortunately, magnetic cycles are much more demanding to detect observationally and only some indication have been seen in few stars. The best known example of a magnetic cycle in another star than our Sun is the very short cycle, much shorter than the solar one, detected in the planet hosting F star $\tau$~Boo (e.g., \cite[Donati \etal\ 2008]{Donati08}). 

The Mt. Wilson H\&K survey provides a unique window to the cyclic activity of 'normal', non-active, stars. The results show that 60\% of the 111 lower main sequence stars studied in the project show cyclic activity (\cite[Baliunas \etal\ 1998]{Baliunas98}). For the 175 studied (sub)giants the fraction is 40\%, and the giants also have a larger fraction of variable activity pattern (not clearly cyclic) than the dwarf stars have (\cite[Baliunas \etal\ 1998]{Baliunas98}).

In active giants the activity cycles are easier to study due to the large brightness variations caused by the huge starspots these stars exhibit. Several studies using long-term photometry have been carried out on the cycles in active giants, e.g., by \cite[Jetsu \etal\ (1990)]{Jetsu90}, \cite[Ol{\'a}h \& Strassmeier (2002)]{Olah02}, and \cite[{\"O}zdarcan \etal\ (2010)]{Ozdarcan10}. Most studies on active giants reveal complex activity cycles with multiple periodicities. A study by \cite[Ol{\'a}h \etal\ (2009)]{Olah09} studied a sample of active stars, among them several giants, using time-frequency analysis. They found that the active stars typically show multiple periodicities, and that the cycle period also changes with time.

In some of the active giants also different kinds of cycles, so-called flip-flop cycles, have been reported. Flip-flops were originally discovered in FK~Com by \cite[Jetsu \etal\ (1993)]{Jetsu93}. In this phenomenon the spot activity concentrates alternatingly on two permanent active longitudes, which are 180$^{\circ}$ apart from each other. The active longitude remains more active for few years, and then the other active longitude takes over and is more active. The change between the active longitudes has been reported to happen in FK~Com every few years (\cite[Korhonen \etal\ 2002]{Korhonen02}). The flip-flop phenomenon has been reported also in few other giant stars, e.g., II~Peg, $\sigma$~Gem, EI~Eri and HR~7275 (\cite[Berdyugina \& Tuominen 1998]{Berdyugina98}). On the other hand, recent results on the flip-flop phenomenon imply that the timing of the phenomenon is not regular (e.g., \cite[Lindborg \etal\ 2011]{Lindborg11}; \cite[Hackman \etal\ 2012]{Hackman12}) and that the change in longitude is often not 180$^{\circ}$ (e.g., \cite[Ol{\'ah \etal\ 2006}]{Olah06}).

\section{Surface differential rotation}

Differential rotation is one of the main ingredients in the dynamo models, which seek to explain the observed magnetic activity in the Sun and other stars. Therefore, it is crucial for understanding the creation of the observed magnetic field to also investigate the differential rotation. 

The theoretical calculations by \cite[Kitchatinov \& R{\"u}diger (1999)]{Kitchatinov99} predict that the differential rotation is larger in the giant stars than in the dwarfs. \cite[Barnes \etal\ (2005)]{Barnes05} measured surface differential rotation for 10 young G2--M2 dwarfs and reported an increase in the magnitude of differential rotation towards earlier spectral types. Similar results have also been obtained by \cite[Saar (2011)]{Saar11}. \cite[K{\"u}ker \& R{\"u}diger (2012)]{Kuker12} use mean field model to study differential rotation of KIC~8366239, a giant star with rotation period of 70 days. The results are compared to the model of Arcturus (\cite[K{\"u}ker \& R{\"u}diger 2011]{Kuker11}), which has a rotation period of the order of two years. The model for Arcturus, with its small core, predicts a steep increase of the rotation rate near the center of the star. On the other hand, the faster rotating KIC~8366239, which also is predicted to have a larger core than Arcturus, has similar internal rotation behaviour as the Sun (not the strength, but the behaviour with radius). Surface differential rotation has been measured for a handful of giant stars. The results collected by \cite[Marsden \etal\ (2007)]{Marsden07} indicate that the active (sub)giants show similar strength of surface differential rotation as the young active stars. More measurements of differential rotation on giant stars are needed to answer the question whether the active giants and dwarfs show similar behaviour.

Anti-solar differential rotation, where the polar regions rotate faster than the equator, has been suggested by observations of several active giants, e.g., in HR~1099 by \cite[Vogt \etal\ (1999)]{Vogt99} and in $\sigma$~Gem by \cite[K{\H o}v{\'a}ri \etal\ (2007)]{Kovari07}. \cite[K{\H o}v{\'a}ri \etal\ (2007)]{Kovari07} studied surface flow patterns on $\sigma$~Gem from observation spanning 3.6 consecutive stellar rotations, and found evidence for a weak anti-solar differential rotation together with indications of poleward migration of spots. This strong meridional flow hinted at $\sigma$~Gem would support the hypothesis of \cite[Kitchatinov \& R{\"u}diger (2004)]{Kitchatinov04}, which attributes the anti-solar differential rotation to strong meridional circulation. Similar trend is implied by the results of \cite[Weber (2007)]{Weber07} on several active giants. One should not forget, though, that these meridional flow measurements can be caused by artifacts in maps and have to be confirmed with data from several epochs.

A recent study by \cite[Korhonen \& Elstner (2011)]{Korhonen11} used snapshots from dynamo simulations to study surface differential rotation obtained using cross-correlation method. The input rotation law could be recovered from the analysis of the model snapshots, but if small-scale fields were included in the models. With using only the large scale dynamo field the input rotation law was not recovered, and usually basically solid body rotation was obtained. This rises the question whether the huge starspot of active giants can actually be created by small scale fields. If they are manifestations of the large-scale dynamo field, then according to the study by \cite[Korhonen \& Elstner (2011)]{Korhonen11} we would not even expect them to follow the surface differential rotation.

\section{Closing remarks}

Rapidly rotating G--K giant stars are among the most magnetically active stars known. They show large starspots, choromospheric and coronal activity, frequent flaring and activity cycles. Slowly rotating G--K giants can also have magnetic fields, maybe through weak dynamo action or due to relic fields (being descendants of Ap stars). In some cases magnetic fields have been detected also in M giants, but no clear explanation for the field generation is known. For the faster rotating ones the field could be generated by a solar-like dynamo. Magnetic field has also been detected in the supergiant Betelgeuze, and it has been hypothesised that a local dynamo can operate in its giant convection cells. 

This all shows that magnetic fields are present also in evolved stars, and in large fraction of them than earlier thought. Still, the creation of these fields is often not well understood -- even the solar dynamo is not fully understood. More observations of magnetism in evolved stars are needed to collect enough clues to unravel its mystery.

{\bf Acknowledgments}
The author acknowledges the support from an IAU travel grant to participate the conference.

\end{document}